\documentstyle[aps,graphicx, times,floats]{revtex}
\graphicspath{{/home/ykao/PAPERS/GLASS/FIGURES/}}
\DeclareGraphicsExtensions{.eps,.ps}
\begin{document}
\draft

\wideabs{

  \title{ History Dependent Phenomena in the Transverse Ising Ferroglass:\\
  the Free Energy Landscape }

  \author{Ying-Jer Kao$^1$, G. S. Grest$^2$, K. Levin$^1$, J. Brooke$^3$,
 T.F. Rosenbaum$^1$ and G. Aeppli$^3$}
 \address{$^1$ James Franck Institute and Department of Physics, University of Chicago, Chicago, Illinois
    60637}
 \address{$^2$Sandia National Laboratory, Albuquerque, NM 87185}
\address{$^3$NEC Research Institute, 4 Independence Way, Princeton, New
 Jersey 08540}
  \date{\today} 

\maketitle
\begin{abstract}
In this paper we investigate the relationship between glassy and
ferromagnetic phases in disordered Ising ferromagnets in the presence of transverse
magnetic fields, $\Gamma$. Iterative mean field simulations probe the free energy landscape
and 
suggest the existence of a glass transition as a function of $\Gamma$ which is distinct from the Curie
temperature.
New experimental field-cooled and zero-field-cooled data on
LiHo$_x$Y$_{1-x}$F$_4$ provide  support for our theoretical picture. Here
as well we present a collection of theoretical predictions for future
experiments.  
\end{abstract}
}

The disordered Ising ferromagnet LiHo$_x$Y$_{1-x}$F$_4$ in a
transverse field $\Gamma$ provides a unique opportunity to study coexistent
glassy and ferromagnetic tendencies. Application of $\Gamma$ tunes both the
glassiness and the ferromagnetism and, thereby, makes it possible to probe
the complex free energy landscape. While the glassy characteristics of the
paramagnetic transverse Ising systems have been studied both
experimentally\cite{Rosenbaum} and 
theoretically\cite{Walasek,Kopec,Ray,Ishii,Goldschmidt,Buttner,dos_Santos,Rieger,Guo},
as has the ferromagnetic transition in the pristine system\cite{Bitko},
little is known about the interplay of ferromagnetism and
glassiness. Equally interesting 
is another unique opportunity afforded by the
LiHo$_x$Y$_{1-x}$F$_4$ system, namely the ability to explore quantum
mechanical effects arising from $\Gamma \ne 0$, in the presence of glassiness\cite{Justin}.

The goal of this paper is to study these two coexistent phenomena
( glassiness and long range magnetic order) via an exploration 
of the free energy
surface, and in this way, make systematic predictions for the 
irreversibility characteristics (i.e., hysteresis, remanent magnetization,
etc.). Ferromagnetism in the presence of significant disorder introduces
a new class of challenging problems.
We show here that the ferromagnetic onset temperature $T_c$ is
distinct from the glass transition temperature $T_g$, and that the latter is
itself dependent on the history of the system. 
Here, in contrast to the rather extensive theoretical studies (of
the paramagnetic phase) in the
literature\cite{Walasek,Kopec,Ray,Ishii,Goldschmidt,Buttner,dos_Santos,Rieger,Guo}, we associated $T_g$, in more physical terms, with the onset of
irreversibility in measurable characteristics. In general, there are three
state variables, including longitudinal fields, $H_z$, as well as $\Gamma$
and $T$, which can be cycled in many different
and non-commuting ways to arrive at
a given minimum in the free energy surfaces.

To support this physical picture, in this paper we also present new experimental data 
showing the difference between
field cooled (FC) and zero field cooled
(ZFC) magnetic \textit{susceptibilities} as a function of $T$ at fixed $\Gamma$ 
in \mbox{LiHo$_{0.44}$Y$_{0.56}$F$_4$}. We find that
this difference vanishes at a (glass transition)
temperature distinct from the Curie point for ferromagnetic ordering.


We base our theoretical analysis on
earlier work in conventional spin glasses\cite{Gary}
and random field systems\cite{Ro} which addressed the
evolution of the free energy landscape using 
an iterative numerical mean field scheme, in which the
reaction terms (which led to problems with numerical convergence\cite{Bowman}) were
ignored.  By following a given free energy minimum as it evolved with field and $T$, 
one arrived at rather good agreement between theory and experiment for the various
history dependent magnetizations. 

This theoretical approach probes the system on intermediate time scales,
which are long compared to the time needed to ``re-equilibrate" after a given free energy
minimum has disappeared (with temperature or field cycling), but short compared to the 
time needed to tunnel\cite{Rosenbaum,Justin} between meta-stable
states on the free energy surface. 

The transverse Ising ferroglass is described by the Hamiltonian:
\begin{equation}
{\mathcal H}=-\frac{1}{2} \sum_{\left<i,j\right>} J_{ij} S_i^zS_j^z - H_z \sum_{i}S_i^z -\Gamma \sum_{i} S_i^x,
\end{equation}
where the sum $\left< ij \right>$ is over the nearest neighbors, 
and the exchange coupling $J_{i,j}$ is given by the Gaussian distribution 
$P(J_{ij})=\sqrt{1/{2\pi J^2}}\exp(-(J_{ij}-J_0)^2/{2 J^2})$,
where $J$ is the variance and $J_0$ is the shift. Here $\Gamma \propto H_t ^2 $
for small $H_t$, where $H_t$ is the transverse magnetic field applied
in the laboratory.
We can obtain a mean-field equation 
 for the average magnetization $m_i ={\left< S_i^z \right>_T}$ ($\left<..\right>_T$ denotes
 thermal averaging) at each site for $S=1/2$:
\begin{equation}
m_i=\frac{{\mathcal P}_i}{4{\mathcal E}_i} \tanh(\beta {\mathcal E}_i)\label{eq:mf},
\end{equation}
where ${\mathcal P}_i=\sum_{<i,j>} J_{ij} m_j+H_z$ and
${\mathcal E}_i=\sqrt{{\mathcal P}_i^2+\Gamma^2}/2$.
This 
mean field equation corresponds to minimizing the
free energy $F$ as a function of the set of $m_i$. 
Here, the transverse field $\Gamma$ effectively enters only
through
the modified Brillouin function of Eq.~(\ref{eq:mf}),
reflecting the fact
that $\Gamma$ rotates the local spin axis away from the Ising or
$z$ direction. Introducing this term in effect mixes the ``$z$-component"
eigenstates of the
$\Gamma = 0$ problem. 
Finally, it should be noted that while the experiments we will present here
address the magnetic susceptibility ($ \chi = \sum \chi_i $
with $\chi_i = \partial m_i / \partial H_z $), 
the calculations are based on the actual magnetizations. Iterative
convergence of the susceptibilities at a given site  
has not yet been established.

\begin{figure}
\centerline{\includegraphics[width=3.2in]{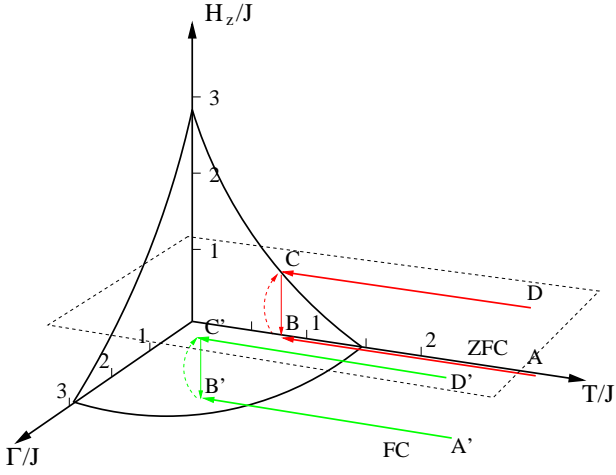}}
\vskip 2mm
\caption{Magnetic field and temperature cycling protocols in the parameter
space used to tune the free energy surface. 
The  ZFC state is achieved through cooling from A to B and FC
  from A' to B'. The remanent magnetizations (called IRM and TRM) are obtained through the paths A-B-C-B,
and D-C-B, respectively. 
The solid lines represent a consolidation of Fig.~\ref{fig:hys} as discussed in
the text below.
}
\label{fig:3d-phase}
\end{figure}
The system in the present study is composed of $N\times N\times N$ spins 
with  random
bond configurations. Most of our examples are for $N=20$ and fixed $J_0$, although we studied
larger ($N=40$)
size systems (and variable $J_0$) to verify convergence of our
results.
In order to make a
closer connection to the experiments\cite{Justin}, we present here the study
of $J_0=0.2J$ so that the system is only slightly ferromagnetic. 
For each set of parameters $(\Gamma,H_z,
T)$, we start our iterations at the ${m_i}$ corresponding to the minimum
of $F$ evaluated at the previous $T, H_z$, or $ \Gamma$. We then update 
${m_i}$ by solving the mean-field equations, Eq.~(\ref{eq:mf}), at each site until 
convergence is obtained at the $n$th iteration defined by 
${\sum_i(m_i^n-m_i^{n-1})^2}/{\sum_i(m_i^n)^2}\le 10^{-6}$.
Unless indicated, the magnetizations discussed in this
paper are taken to be along the $z$- (Ising) axis.
Finally, a small $H_z = 0.01 J$, was applied in all studies of
glassy properties. This was needed to establish a fixed direction for
spontaneous broken symmetry. 

Figure~\ref{fig:3d-phase} is a three dimensional plot of the parameter space which
we consider; these parameters cause the free energy surface to evolve in distinct
ways. Various 
pathways in this parameter space will be referred to throughout the text. The solid
lines shown here are irreversibility contours derived from magnetic hysteresis curves 
that are
discussed later in the paper.

\begin{figure}
\centerline{\includegraphics[width=3.4in]{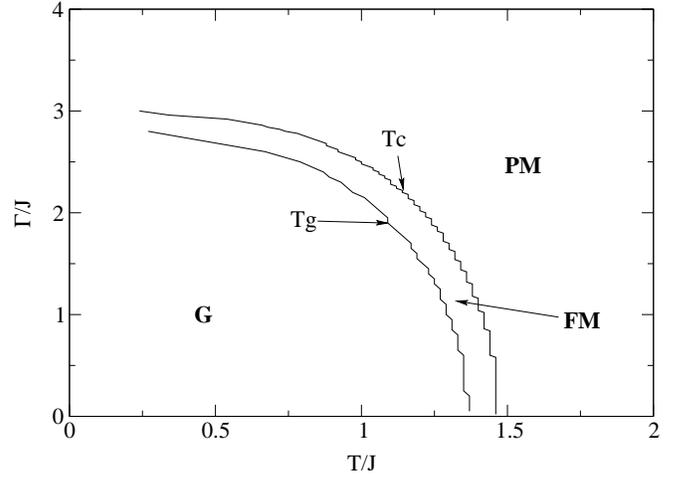}}
\vskip 2mm
\caption{Mean field phase diagram of the Ising ferroglass in
  transverse magnetic field $\Gamma$. Here PM represents paramagnet; FM, ferromagnet; and
  G, glassy state.  $T_g$ is determined by
the point at which the FC and ZFC magnetizations are equal, and, consequently,
is defined
only for finite $\Gamma$. 
}
\label{fig:phase}
\end{figure}

In Fig.~\ref{fig:phase} we present an $H_z = 0$ phase diagram for the Hamiltonian
of Eq.(1), 
omitting the lowest T regime where tunneling effects are important.
There are
  two distinct lines separating different phases: the outer line indicates the phase
  boundary for the paramagnetic to ferromagnetic phases and the 
  inner line is the ferromagnet to ferroglass
  phase boundary. The glassy boundary was determined by the condition that
the transverse field cooled (FC) and zero field cooled (ZFC) magnetizations are
equivalent. The various cooling pathways can be seen in Fig.~\ref{fig:3d-phase}. 
It should be stressed that in vanishing transverse field the quantity $T_g$ has
meaning only as an asymptote, since the distinction between FC and ZFC becomes
meaningless when there is no field.
The para-ferro boundary is defined by the set of $(\Gamma,
  T)$ at which a spontaneous magnetization appears.
 For a given $N$, this magnetization may be calibrated by first establishing
a baseline zero, which is the magnitude of 
``spontaneous magnetization" due to  finite size effects, estimated from the
$J_0 = 0$ case.  
  By changing $J_0$,
  we were able to change the position of the para-ferromagnetic line
 relative to the glassy line; for smaller $J_0$, the para-ferromagnetic
 line will be inside the glassy line. 

We now turn to experiments.
LiHoF$_4$ is a three-dimensional, dipolar-coupled Ising magnet. In the
classical limit ( $\Gamma=0$), the Ho dipoles order ferromagnetically at a
Curie temperature $T_c=1.53$K. 
Experiments confirm that standard 
mean field theory describes fully the
critical behavior of the phase boundary between paramagnet and
ordered ferromagnet\cite{Bitko}.
Moreover, magnetically inert yttrium can be substituted for the holmium spins in
single crystals of LiHo$_x$Y$_{1-x}$F$_4$, permitting carefully controlled
studies of the effects of quenched disorder. We suspended a needle of
LiHo$_{0.44}$Y$_{0.56}$F$_4$ with $T_c(\Gamma=0)=670$~mK from the mixing
chamber of a helium dilution refrigerator into the bore of an 8~T
superconducting magnet aligned perpendicular to the Ising c-axis (within
$0.5^\circ$). A trim coil oriented along the Ising axis nulled stray
longitudinal fields from the magnets. We measured the AC magnetic
susceptibility $\chi'$ in the frequency and excitation independent limits
using a digital lock-in technique. 

\begin{figure}
\centerline{\includegraphics[width=3.4in]{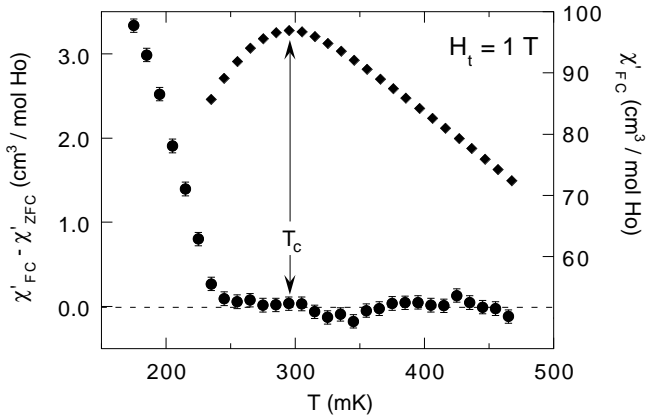}}
\vskip 2mm
\caption{Temperature evolution of the AC magnetic susceptibility $\chi'$ of 
LiHo$_{0.44}$Y$_{0.56}$F$_4$ at $H_t=1$~T for the FC protocol and the difference
in $\chi'$ under FC and ZFC protocols. The Curie temperature($T_c$)  in this
transverse field is $295$~mK while the FC and ZFC susceptibilities bifurcate
at $T=240$~mK.}
\label{fig:exp}
\vspace*{ -0.2in} 
\end{figure}

We compare in Fig.~\ref{fig:exp} the system response after various
trajectories in $H_t-T$ space, where the static longitudinal field has
been carefully set to zero. In the FC protocol, the sample was cooled in
$H_t=1$~T from the paramagnet into the ordered state at $T=175$~mK and
then warmed; in the ZFC protocol, $H_t$ was only applied after cooling to
175~mK. The ferromagnetic transition is marked by a peak in field-cooled
susceptibility at $T=295$~mK. By contrast, the FC and ZFC susceptibilities
bifurcate deeper in the ordered state, at $T=240$~mK.  As is
consistent with the
theoretical mean field phase diagram of Fig.~\ref{fig:phase}, we find
clearly distinct signatures for ferromagnetism and glassiness in the
disordered Ising magnet LiHo$_{0.44}$Y$_{0.56}$F$_4$.

\begin{figure}
\centerline{\includegraphics[width=3.6in]{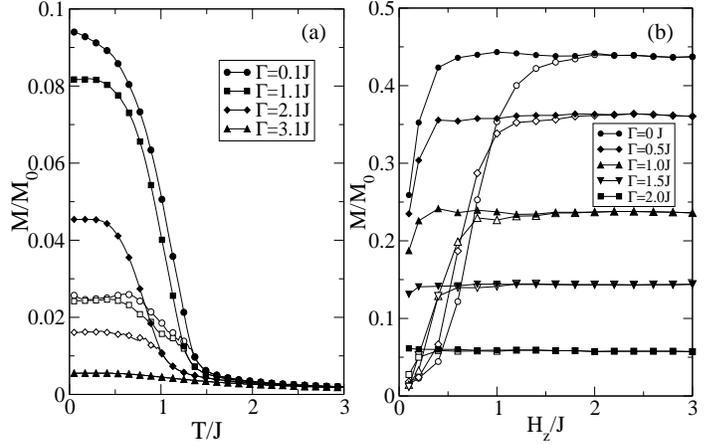}}
\vskip 2mm
\caption{(a) Temperature dependence of zero-field-cooled (ZFC, open symbols) and
  field-cooled (FC, closed symbols) magnetizations for the Ising ferroglass
 at different
  $\Gamma$
(b) Field dependence of IRM(open symbols) and TRM(closed symbols) versus $H_z$ for different $\Gamma$ at
  $T=0.15J$.
}
\label{fig:zfcfc}
\vspace*{ -0.2in}
\end{figure}

To elucidate these data, FC and ZFC processes were simulated numerically 
as in classical spin glasses\cite{Gary}, with the fields applied
in the transverse direction. 
  We first cooled
  down the system in zero external field (path $A \rightarrow B$ in
  Fig.~\ref{fig:3d-phase}). We then applied a transverse
  field, and warmed up in temperature. In this way 
  we obtained  a zero-field-cooled magnetization $M^{ZFC}$ as a function
of  temperature. These results are shown by the open symbols in
Fig.~\ref{fig:zfcfc}(a). We
  next cooled the system in the presence of a transverse field to
$T=0.1J$ (path $A'\rightarrow B'$)
and then warmed up in the presence of this
  field. We thereby obtained the $M^{FC}$
  curve (indicated by the solid symbols in Fig.~\ref{fig:zfcfc}(a)). 
The two magnetizations merge at a given 
  temperature $T_g$, which we identify as the glass transition
  temperature of Fig.~\ref{fig:phase}. We found that $M^{FC}$ is  reversible upon
  cooling and warming, while $M^{ZFC}$ is not reversible for a subsequent
  cooling from a temperature lower than $T_g$. 
In this way we can regard the $FC$ state as that which is closer to
true thermodynamical equilibrium, and it should not be surprising
that this state has the larger magnetization of the two; this larger
$M$ takes advantage of the net ferromagnetic bias $J_0$. 
As the system is warmed
the magnetization decreases monotonically.
For both
  $M^{FC}$ and $M^{ZFC}$, the magnetizations at the lowest temperatures
  decrease as $\Gamma$ increases. This is a consequence
of the off-diagonal components introduced by
  $\Gamma$, which act to reduce the net
  magnetization in the $z$ direction.

We turn now to the issue of remanent magnetizations. These remanences arise
following the removal of a magnetic field, when the system becomes
trapped in a metastable minimum. In canonical Ising spin
glasses there are two characteristic remanences\cite{Gary} (associated with $H_z$).
These are the isothermal remanent magnetization (IRM) and
the thermal remanent magnetization (TRM).
In the presence of a \textit{fixed} transverse field we modify their definitions slightly, as follows.
We first consider the case $\Gamma = 0$.
When the system is cooled through path $A\rightarrow B$ in Fig.~\ref{fig:3d-phase} and
a longitudinal field is applied instantaneously ($B\rightarrow C$) and subsequently
adiabatically
removed ($C\rightarrow B$), we refer to the resulting magnetization
as the thermal remanent magnetization. When
the system is cooled through path $D\rightarrow C$ in a longitudinal field and the
field is adiabatically turned off ($C\rightarrow B$), we refer to the
resulting magnetization as the isothermal remanent
magnetization. For the present purposes we will be interested in
the IRM and TRM at finite $\Gamma$.  This corresponds
to the paths $A'\rightarrow
B'\rightarrow C'\rightarrow B'$ and $D'\rightarrow B'\rightarrow C'$ respectively. 
The behavior of the two remanences at different $\Gamma$ is shown in Fig.~\ref{fig:zfcfc}(b) for
$T=0.15J$. Both the TRM (closed symbols) and the IRM (open symbols) become smaller 
as $\Gamma$
increases, as expected since the spins are aligned more toward the transverse
field direction.
We find that
the $H_z$ value where the two remanences are equal 
becomes smaller as $\Gamma$ increases, indicating that at higher $\Gamma$ it
takes less $H_z$ to destroy the multiple minima on the free energy surface. 

\begin{figure}
\centerline{\includegraphics[width=3.6in]{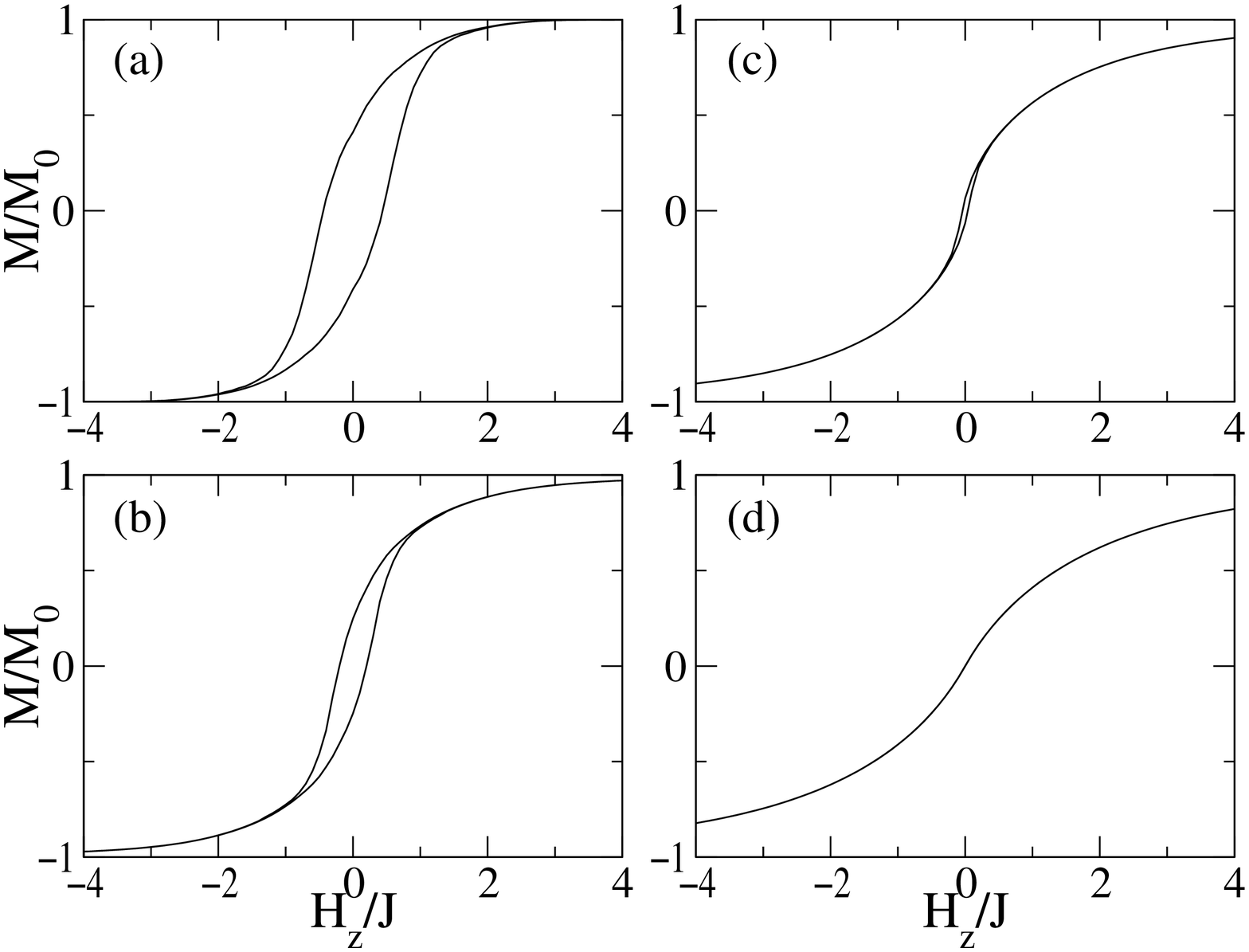}}
\vskip 2mm
\caption{Hysteresis loops at fixed $\Gamma=$: (a)$0.0J$, (b)$1.0J$,
  (c)$2.0J$ and (d)$3.0J$. Here $T=0.15J$.}
\label{fig:hys}
\end{figure}

As a final protocol, we address the behavior of magnetic hysteresis loops (for the
magnetization as a function of varying $H_z$),
in the presence of
 fixed transverse fields. The results are shown in
  Figs.~\ref{fig:hys}(a)-(d), where the transverse fields are $\Gamma=0.0J, 1.0J,
  2.0J $ and $3.0J$ respectively. The loops were obtained by slowly
  decreasing the parallel
  field ($H_z$) from the high field limit ($4J$) until an equally large negative
  field was reached and then sweeping back to close the loop. 
  As is evident,
a non-zero transverse field enables the hysteresis loop to close at smaller $H_z$.

We may summarize
these numerical hysteresis studies in a three dimensional
plot of the associated irreversibility surface 
(solid lines in Fig.~\ref{fig:3d-phase}).
The surface consists of the locus of points in terms of the coordinates
$(\Gamma, H_z, T)$ 
below which the system exhibits irreversibility, as reflected in magnetic
hysteresis loops.
It should be stressed that this protocol is different from that used to
obtain $T_g$ in Fig.~\ref{fig:3d-phase}. 
While both protocols are experimentally accessible, the point at
which irreversibility sets in for a given $( \Gamma, H_z, T )$
is not unique, and, itself, depends upon the pathway to the point
in question.  This scenario is to be distinguished from the more
conventional situation which in which either $\Gamma$ or
$H_z$ is strictly set to zero, as in Refs~\onlinecite{Gary}
and Refs~\onlinecite{Walasek} - \onlinecite{Guo}, respectively.
An important conclusion from Fig.~\ref{fig:3d-phase} 
is that raising the temperature
or applying a longitudinal or transverse magnetic field progressively
removes minima from the free energy surface until at sufficiently high
temperatures or fields there is a unique state.

The calculations which underlie this figure
ignore thermal, as well as quantum, fluctuations.  Given the latter,
the low temperature regime should be viewed as inaccessible.
Our results can be contrasted with a recent low temperature study (in the paramagnetic
phase) which investigated glassy phase in a quantum $p$-spin spherical model 
within an equilibrium statistical mechanical
approach\cite{Cugliandolo}. Here it was observed that hysteretic effects may 
also arise from a first-order phase transition, rather than from the glassiness
which we have emphasized here. It should be noted that the experimental
hysteresis reported in Fig.~\ref{fig:exp} is not strictly in the quantum regime. However,
to make a firm distinction between these two theoretical
scenarios will require further experiments.  

In summary, in this paper we have emphasized the concept of
history dependent measurements in transverse Ising ferroglasses,
a concept that has been widely recognized in other magnetic glasses\cite{Birgeneau}. 
Because of its focus on the complex free energy landscape, our approach should be contrasted 
with alternatives in the literature,
which have also addressed the phase diagram of LiHo$_x$Y${_{1-x}}$F$_4$,
but with an emphasis on the low temperature quantum regime using equilibrium
statistical mechanics. It should also be stressed that these transverse field 
configurations represent a unique opportunity
to simultaneously tune glassiness along with long range ferromagnetic
order.  Our predictions for the \textit{intermediate
time scale} behavior 
of the various history
dependent magnetizations
appear consistent with the FC and ZFC
susceptibility measurements presented here, but further experiments
will be needed to confirm the sets of the predictions presented here.
      
We thank Qimiao Si for useful conversations. 
The work at the University of Chicago was supported by the NSF-MRSEC, under award No. DMR-9808595.
Sandia is a multiprogram
laboratory operated by Sandia Corporation, a Lockheed Martin Company, 
for the United States Department of Energy under Contract DE-AC04-94AL85000.



\end{document}